# Effect of Ga doping on magneto – transport properties in collosal magnetoresistive $La_{0.7}Ca_{0.3}Mn_{1-x}Ga_xO_3$ (0 < x < 0.1)

Marek Pękała
Department of Chemistry, University of Warsaw
Al. Zwirki i Wigury 101, PL-02-089 Warsaw, Poland,

Jan Mucha
Institute of Low Temperature and Structural Research
Polish Academy of Sciences, PL-50-950 Wroclaw, Poland

Benedicte Vertruyen and Rudi Cloots
S.U.P.R.A.T.E.C.S., University of Liege, Institute of Chemistry B6
Sart Tilman, B-4000 LIEGE Belgium

Marcel Ausloos
S.U.P.R.A.T.E.C.S., University of Liege, Institute of Physics B5
Sart Tilman, B-4000 LIEGE Belgium

**ABSTRACT**
Samples of $La_{0.7}Ca_{0.3}Mn_{1-x}Ga_xO_3$ with x = 0, 0.025, 0.05 and 0.10 were prepared by standard solid-state reaction. They were first characterized chemically, including the microstructure. The magnetic properties and various transport properties, i.e. the electrical resistivity, magnetoresistivity (for a field below 8T), thermoelectric power and thermal conductivity measured each time on the same sample, are reported. The markedly different behavior of the x = 0.1 sample from those with a smaller Ga content, is discussed. The dilution of the $Mn^{3+}/Mn^{4+}$ interactions with Ga doping considerably reduces the ferromagnetic double exchange interaction within the manganese lattice leading to a decrease of the Curie temperature. The polaron binding energy varies from 224 to 243 meV with increased Ga doping.

## 1. INTRODUCTION

Since the early nineties, there has been a renewed interest for the mixed-valence manganites $Ln_{1-x}A_xMnO_3$ (where Ln is a lanthanide cation and A is usually an alkaline earth ion), due to the colossal magnetoresistance (CMR) properties displayed by some compounds of that family [1,2]. On one hand, films [3] of these materials have potential applications, e.g. as magnetic sensors or in computer memory systems [4]. On the other hand, it is expected that the understanding of the underlying physical mechanisms could lead to substantial advances in the field of strongly correlated electron physics.

CMR compounds such as $La_{0.7}Ca_{0.3}MnO_3$ are characterized by a maximum in magneto - resistance at a coupled to a transition involving electrical and magnetic properties, i.e. the system goes from a ferromagnetic/metallic-like phase at low temperature to a paramagnetic/insulating phase at higher temperature. Since many different cations can be substituted either on the Ln or on the Mn site, a large amount of experimental data is now available [5]. It turns out that the various exchange interactions in the three-dimensional manganese oxygen network imply considerable variability of structures. E.g. recently [6,7] we have substituted a small amount of Y on the Ln site and discovered a rich phenomenology in the specific heat and electrical resistivity of $La_{0.6}Y_{0.1}Ca_{0.3}MnO_3$ in zero magnetic field due to short range magnetic ordering in disordered clusters, in particular due to the Mn-O-Mn bond environment. Magnon-polaron and spin-polaron signatures were revealed together with charge localization mechanisms. Introducing foreign nonmagnetic cations on the Mn site of mixed - valency manganites $Ln_{1-x}A_xMnO_3$ is also known to lead to severe modifications of the magnetic and electrical properties, as well in conductive ferromagnetic manganites [8-10] as in insulating charge-ordered compounds [11,12]. ]. In the ferromagnetic region due to Mn ions replacement with such metals there are sometimes features ascribed to a weakening of the Zener double-exchange interaction between two unlike ions [13]. It seems that more data, and on different properties that the electrical resistivity could bring some further information on the microscopic mechanisms.

In the present work we have studied the substitution of a small amount of gallium ions onto the manganese site of $La_{0.7}Ca_{0.3}MnO_3$. The $Ga^{3+}$ ion has the same charge and a similar ionic size as $Mn^{3+}$, but has no magnetic moment ($d^{10}$ configuration). Thus Ga doping does not cause lattice distortions. Due to these specific characteristics, Ga-substituted manganites have often been used as a model system for studying super-exchange interactions in $LaMnO_3$ (see [14-22]). Meera et al. [18] and Yusuf et al. [19] studied the magnetic properties of $La_{0.67}Ca_{0.33}Mn_{1-y}Ga_yO_3$ samples with y = 0.1. Cao et al. [20] reported X-ray absorption results for $La_{0.7}Ca_{0.3}Mn_{1-y}Ga_yO_3$ compounds with y ≤ 0.1. Sun et al. [21] published a paper focused on the paramagnetic regime of $La_{0.67}Ca_{0.33}Mn_{1-y}Ga_yO_3$ samples with y ≤ 0.1. An additional interest in the gallium doped manganites flows from a possible quantum critical point predicted for such a material [22].

This paper was intended to obtain the comprehensive experimental characterization of the $La_{0.7}Ca_{0.3}Mn_{1-y}Ga_yO_3$ systems, including the novel results of thermal conductivity measurements. The systematic investigations of all the magnetic and transport properties were performed on the same and well defined samples. Here below the results of magneto - transport investigations, i.e. of the electrical resistivity, thermoelectric power and thermal conductivity, are presented and discussed. In Sect. 2, we briefly outline the experimental techniques. In Sect. 3 the samples are characterized with respect to their chemical content and microstructure. In Sect. 4 the magnetic properties are given. Sections 5-9 contain results and comments on the various transport properties, measured on the same sample in magnetic field up to 8 T. A brief discussion and a conclusion are found in Sect. 10.

## 2. EXPERIMENTAL

Samples of $La_{0.7}Ca_{0.3}Mn_{1-x}Ga_xO_3$ with x = 0, 0.025, 0.05 and 0.10 were prepared by standard solid-state reaction. Stoichiometric amounts of pre-calcined lanthanum oxide ($La_2O_3$), calcium carbonate ($CaCO_3$), manganese oxalate ($MnC_2O_4.2H_2O$) and gallium oxide ($Ga_2O_3$) were thoroughly ground in petroleum ether and heated up to 1200°C with intermediate grindings. Pellets were pressed uniaxially and sintered at 1250°C during 48 h (samples with x = 0 and 0.025) or 96 h (samples with x = 0.05 and 0.10).

X-ray diffraction patterns were collected with a Siemens D5000 diffractometer ($CuK_\alpha$ radiation). The microstructural properties were characterized by scanning electron microscopy (Philips XL30 ESEM) and optical microscopy with polarized light (Olympus AH3-UMA). The cationic composition was checked by energy dispersive X-ray analysis (EDX) coupled to the electron microscope.

The magnetic characterization was performed in a Physical Property Measurement System (PPMS) from Quantum Design. The m(H) curves were collected at 10 K. The AC susceptibility was measured as a function of temperature for a 10 mT and 1053 Hz applied AC magnetic field.

The electrical resistivity was measured by the four probe technique using currents at most 10 mA. Thermoelectric power and thermal conductivity were simultaneously measured at zero field and under a 8 T field on the same sample using the differential and steady flow methods with a temperature gradient about 1 K created by a mW heater. In order to **minimize** radiation effects on thermal conductivity the radiation shield around samples was kept at the sample temperature. Moreover, thermal conductivity was checked to be independent of applied temperature gradient. **Inaccuracy of thermal conductivity due to radiation losses and wire thermal conductivity is about 1 %, whereas absolute values are determined with 3 % error mainly due to irregular sample geometry.** The magneto - transport measurements were basically similar to those described previously [23,24].

## 3. SAMPLE CHARACTERISATION

The X-ray diffraction patterns of all samples were seen to display the characteristic peaks of a single-phase perovskite manganite structure and were indexed in the Pnma space group. The EDX analysis showed no deviation from the nominal cationic composition within the limit of the technique accuracy. In the case of the sample with $x = 0.10$, a few grains of some oxide of gallium and manganese were detected.

From polarized light microscopy on sample cuts we observed that the grain cross sections are irregular, varying between 4 and 16 $\mu m^2$. Grains are relatively smaller for the samples with $x = 0$ and 0.025, and larger for the samples with $x = 0.05$ and 0.10, which were sintered during a longer time, as recalled in section 2.

## 4. MAGNETIC PROPERTIES

Figure 1 shows the curves of the DC magnetization at 10 K as a function of the magnetic field up to $\mu_0H = 3T$. All samples display a ferromagnetic behaviour, with coercive fields less than 10 mT. In the inset of figure 1, the experimental values of the saturation magnetic moment per formula unit at 3 T and 10 K (black symbols) are compared to the theoretical "spin-only" magnetic moments (dashed line). The theoretical "spin-only" magnetic moments are calculated assuming that magnetic moments per $Mn^{3+}$ and $Mn^{4+}$ ions equal to 4 $\mu_B$ and 3 $\mu_B$, respectively. When fractions of the $Mn^{3+}$ and $Mn^{4+}$ ions are taken as ($0.7 - x$) and 0.3, respectively, the calculated magnetic moment diminishes proportionally to x. Differences between the experimental and theoretical magnetic moments do not exceed 0.01 $\mu_B$. This proves that the "spin only" values are adequate at this Ga doping level. Moreover, this shows that practically all manganese ions contribute to the ferromagnetic phase. No antiferromagnetic ordering has to be assumed.

The temperature dependence of the in-phase component ($\chi'$) of the AC magnetic susceptibility (for 1 mT and 1053 Hz) is shown in Figure 2. The temperature of the magnetic transition decreases when the gallium content increases. The Curie temperatures $T_C$ (at the largest slope of the $\chi'(T)$) are reported in Table 1. For x between 0 and 0.05, the curves display a typical ferromagnetic behaviour, with only a very smooth decrease of the magnetic susceptibility below $T_C$. On the contrary, for the sample with $x = 0.10$, the magnetic susceptibility drops significantly below ~ 90K. The out-of-phase component ($\chi''$) of the AC magnetic susceptibility of this sample is also shown in Figure 2. This curve presents two maxima, at temperatures of 116 K and 86 K, corresponding respectively to the ferromagnetic transition and the decrease of magnetic moment observed in the $\chi'(T)$ curve.

The temperature dependencies of the DC magnetization under 10 and 100 mT for the sample with $x = 0.10$ are shown in Figure 3. Just below the Curie temperature one may distinguish the pronounced split between the field-cooled (FC) and zero-field-cooled (ZFC) curves for $\mu_0H = 10$ mT. On the contrary, the transition interval is broadened and there is only a small splitting below 40 K between the field-cooled (FC) and zero-field-cooled (ZFC) curves for $\mu_0H = 100$ mT. Such a weak irreversibility behavior suggests that the magnetic anisotropy is not large in the $La_{0.7}Ca_{0.3}Mn_{0.9}Ga_{0.1}O_3$ sample. In the range where $\chi'(T)$ drops rapidly, no specific feature can be observed in magnetization below 90 K,. The Curie temperatures shifting gradually from 249 K for $x = 0$ towards 122 K for $x = 0.1$, clearly show that the Ga doping weakens the double exchange coupling among the manganese ions.

## 5. ELECTRICAL RESISTIVITY

Each measured electrical resistivity (Fig. 4 and 5) of the $La_{0.7}Ca_{0.3}Mn_{1-x}Ga_xO_3$ systems follows the characteristic pattern of magnetoresistive materials, exhibiting the two phase behavior with the metallic and semiconducting variations separated by pronounced maximum at the $T_P$ (peak) temperature. For all compositions studied the electrical resistance maximum shifts towards high temperatures approximately proportionally to the applied magnetic field, as shown in Fig. 6. This $T_P$ shift is accompanied by a remarkable reduction of the resistance maximum magnitude.

At low temperatures the absolute values of the electrical resistivity are only weakly affected by the applied magnetic field. For the zero field case electrical resistivities at 10 K are about 2 to 10 x $10^{-5}$ $\Omega$m for x = 0 to 0.05 (Tab. 1). The electrical resistivity magnitude agrees with values reported by Sun et al. [21]. A considerably higher electrical resistivity value, i.e. about 0.2 $\Omega$m, is found for the x = 0.1 sample. The relatively high values of the ratio of the electrical resistivities at room and at low temperature ( $R_{RT}$ / R(10K) ) reach up to 57, proving a very good intergrain connectivity and more generally the high quality of the samples.

In the metallic phase the measured electrical resistivity between 10 K and $T_P$ was fitted to the power type expression [25,26]

$$\rho(T) = \rho_0 + A\,T^N \quad (1)$$

where $\rho_0$ is a resistivity at 10 K, value given in Table 1. The values of the N exponent are plotted in Fig. 7. They increase from about 3 to 6 when passing from x = 0 to x = 0.1. Values of N diminish gradually with increasing magnetic field.

In the semiconducting phase, between $T_P$ and room temperature, thermally activated transport processes described by

$$\rho(T) = A\,T\,\exp(E_R/kT) \quad (2)$$

are observed from semi-log plots. The derived activation energy $E_R$ increases from 117 to 151 meV when x varies from 0 to 0.1 (Fig. 8). .. For x below 0.1, the values of $E_R$ diminish with the magnetic field. This proves that the external magnetic field suppresses the transport energy barriers. In other words, similar to the effects of an applied magnetic field [6], this metal-ion doping is again found [13] to decrease the polaron tunneling energy barrier (thus increasing the correlation length).

A different behavior is observed for the sample with x = 0.1, which has relatively higher electrical resistivity. This higher electrical resistivity for x = 0.1 conforms to the tendency, which would consist in enhancing the electron localization with increased Ga doping [20-22], as discussed later. Values of $E_A$ attain a minimum at 4 T and then increase up to 156 meV at 8 T.

## 6. MAGNETORESISTANCE

All the samples studied exhibit a strong negative magneto-resistance (MR) effect (Fig. 9 ). The most abrupt increase of MR is observed at low magnetic fields, when the MR sensitivity reaches even 11 % per 1 T for x = 0.1. At the highest applied (8 T) magnetic field the MR peak exceeds 92 % for all the samples studied. The magnitude of MR only weakly depends on Ga doping, in contrast to the reports by Cao et al. [20]. The temperature position of the (relatively sharp) magnetoresistance maximum is located close to the $T_P$ temperature and seems to be only weakly affected by the magnetic field.

In order to evaluate an influence of magnetic field on magneto - resistance, the area over the MR curves of Fig 9, was integrated in the same temperature interval for various magnetic fields. It is found for all the samples studied that the MR area varies proportionally to $H^n$ with n equal to 0.58 ± 0.03.

The correlation coefficients R of the electrical resistivity was fitted to eq. 2, listed in Table 2, show that for $0 \le x \le 0.05$ the agreement between the experimental results and the adiabatic polaron model weakens with increasing magnetic field. An opposite tendency in R variation is found for the manganite with x = 0.1, which exhibits the high resistivity. The reason might be due to an increasing short range ordering in the latter compound.

## 7. THERMOELECTRIC POWER

Thermoelectric power of $La_{0.7}Ca_{0.3}Mn_{1-y}Ga_yO_3$ samples is negative in the whole temperature range studied between 30 and 300 K (Fig. 10). For the $La_{0.7}Ca_{0.3}MnO_3$ manganite the thermoelectric power is a smooth function of temperature exhibiting a broad minimum of -30 µV/K around 200 K. The sign and temperature variation of TEP agree with those reported for the $La_{0.7}Ca_{0.3}MnO_3$ systems in [27,28]. The present measurement shows no trace of the TEP minimum around 25 K reported in earlier studies [25].

The Ga doping on Mn sites remarkably affects the temperature variation and the magnitude of TEP. There is much similarity between the behaviors of the TEP as a function of temperature at different Ga doping. The TEP is negative and decreases smoothly up to a $T_B$ break point (Table 1). For x < 0.1 the TEP break point is located about 10 to 20 K below the electrical resistivity maximum in zero magnetic field (Fig. 5). It can be observed for the Ga doped samples, that after the break the variation of TEP is quasi linear with a slope = ca. -5/3 µV/K$^2$ (practically the same for all Ga doped samples) and attains a field dependent minimum (Table 1). The TEP minimum shifts towards lower temperatures and lower values with higher Ga content (Fig. 10). Then TEP rises smoothly and all TEP curves seem to merge at a -20 µV/K value at a temperature ca. 340 K, as indicated by a smooth extrapolation. The present results reveal the high sensitivity of TEP to the Ga doping in the lanthanum-calcium manganites. It is worth to notice that the negative TEP of the manganites studied is in contrast to the low and positive TEP found for the manganites with very close composition $La_{0.67}Ca_{0.33}Mn_{1-y}Ga_yO_3$ (y = 0.02, 0.05, 0.1) [21]. On the other hand, one may remind that the manganites studied have a composition very close to this one ( 0.25 < x < 0.33 ) corresponding to the TEP sign change reported by [30]. According to the formula as eq.2, the activation energy $E_Q$ derived from TEP vs temperature variation in the vicinity of the room temperature lies in a range from 5 to 40 meV, when passing from zero to 0.1 Ga content.

## 8. THERMAL CONDUCTIVITY

The thermal conductivity k(T) of the $La_{0.7}Ca_{0.3}MnO_3$ and $La_{0.7}Ca_{0.3}Mn_{0.95}Ga_{0.05}O_3$ samples are of the same order of magnitude (Fig. 11). The magnitude and temperature variation for $La_{0.7}Ca_{0.3}MnO_3$ are very close to previous reports for $La_{0.7}Ca_{0.3}MnO_3$ [31]. In zero magnetic field both the undoped sample and the one containing 5 % of Ga exhibit a maximum k(T) around 30 K, i.e. a usual characteristic for metals and dielectrics. The applied magnetic field shifts these maxima towards somewhat higher temperatures temperatures than in the case of the electrical resistivity. The causes are thought to arise from a competition between the anharmonic scattering of heat carriers (predominant at high temperature) and the elastic scattering on defects (predominant at low temperature), grain boundaries and sample boundaries. A very similar thermal conductivity maximum was also reported for $La_{0.7}Ca_{0.3}MnO_3$ in [32,33].

Above its maximum the thermal conductivity of $La_{0.7}Ca_{0.3}MnO_3$ can be approximated by $k(T) \sim T^{-0.37}$ up to 100 K. For the $La_{0.7}Ca_{0.3}Mn_{0.95}Ga_{0.05}O_3$ sample a similar formula applies only up to 70 K. Since the exponent is lower than 1, a value expected for purely anharmonic effects, we may conclude that some additional mechanisms, beyond those listed here above, play a role in such materials. In zero magnetic field the thermal conductivity of $La_{0.7}Ca_{0.3}Mn_{0.95}Ga_{0.05}O_3$ samples diminishes monotonically down to a minimum at 240 K - about 20 K below the electrical resistivity maximum. Above this minimum k(T) varies almost proportionally to temperature. This removes any noticeably spurious effect of radiation, which should be proportional to $T^3$. Unfortunately now it is impossible to distinguish a specific mechanism responsible or dominating in this temperature range before performing further studies.

The temperature variation of the thermal conductivity of $La_{0.7}Ca_{0.3}Mn_{0.95}Ga_{0.05}O_3$ is more complicated exhibiting a maximum around 150 K. The mechanism leading to this second maximum is not yet fully understood. The electronic contribution to k(T) estimated from electrical resistivity by a Wiedemann – Franz relation, does not exceed 1 % of measured k(T) and proves that heat is transported overwhelmingly by phonons. The phonon contribution to k(T) varies as $T^{-1}$ due to anharmonic mechanism [34]. On the other hand, the magnon one depends on temperature as $T^2$, as shown by C. Hess *et al.* [32,33] for $La_2CuO_4$. Thus, the second k(T) maximum can be thought to result from a superposition of the phonon and magnon conductivities, exhibiting opposite temperature dependencies. Notice however that the electronic scattering by magnetic impurities can also lead to a peak in k(T) at moderate temperature [35]. For the $La_{0.7}Ca_{0.3}Mn_{0.95}Ga_{0.05}O_3$ sample the k(T) minimum is located at 170 for zero magnetic field. Also for this sample the k(T) minimum coincides with a resistivity maximum for zero magnetic field. The radiation effects are not seen in the roughly linear variation of k(T) above the minimum.

## 9. MAGNETOTHERMAL CONDUCTIVITY

An application of a magnetic field of 4 T, removes the k(T) minimum found in zero magnetic field for the thermal conductivity of $La_{0.7}Ca_{0.3}MnO_3$ samples (Fig. 11a) around 240 K - about 20 K below the electrical resistivity maximum. When applying a 4 T magnetic field, the residual k(T) minimum is shifted up to 280 K, only 10 K below the electrical resistivity maximum (Fig. 5A). An influence of external magnetic field on heat transport is further confirmed for $La_{0.7}Ca_{0.3}Mn_{0.95}Ga_{0.05}O_3$ system by a comparison of k(T) in zero and 8 T fields (Fig. 11b). In the 8 T magnetic field k(T) minimum shifts to 200 K - about 40 K below the electrical resistivity maximum (Fig. 5C).

The role of an external magnetic field is additionally demonstrated in Fig. 12 showing the difference in thermal conductivity D(T) = k(T, H) - k(T, H=0). For $La_{0.7}Ca_{0.3}MnO_3$ sample the D(T) curve has a broad maximum around 240 K, which coincides with the electrical resistivity maximum. For a sample containing 5 % of Ga the D(T) maximum appears at 150 K and the values of D(T) rise abruptly to above 220 K.

## 10. DISCUSSION

The $Ga^{3+}$ ion has a complete d-shell, whence does not have any magnetic moment. The very good agreement demonstrated between the experimental values for the saturation magnetization and the theoretical spin-only values shows that gallium substitution up to x = 0.10 does not prevent the long range ferromagnetic alignment of the $Mn^{3+}$ and $Mn^{4+}$ ions. The neutron - diffraction studies have shown that replacement of Mn by Ga induces a continuous decrease of the tetragonal distortion of the $(Mn/Ga)O_6$ octahedra, which become practically regular for x = 0.6 [36]. On the other hand, the dilution of the $Mn^{3+}/Mn^{4+}$ interactions by the gallium ions do not contribute to ferromagnetic exchange leading to a decrease of the Curie temperature.

For x up to 0.05, there is a good agreement between $T_C$ and the temperature of the resistance maximum in the R(T) curves. On the contrary, for x = 0.10 there is a difference of ~50 K between $T_C$ (129 K) and the resistance maximum (82 K). However, the MR(T) curves for the sample with x = 0.10 display a shoulder in the temperature range near $T_C$. The Ga doping affects also the transport processes in manganites. The lattice sites occupied by the $Ga^{3+}$ ions constitute the barriers for the conduction carriers, which in turn reduces a number of the active conduction paths. The EDX analysis did not reveal any noticeable diferences in Ga density at grain boundaries and in grain cores of manganites studied. Such an homegeneous distribution og Ga is in contrast to this compounds doped with Ru transtion metal. It was observed for the Ru doped manganites that the enhanced dopant density at grain boundaries leads to the additional electron scattering processes at grain boundaries [37,38].

Applying the polaron model and combining the activation energies $E_R$ and $E_Q$ determined from the electrical resistivity and thermoelectric power data, respectively, the polaron binding energy $E_B$ was determined for manganite samples with various Ga content according to a formula [21]

$$E_B = 2 ( E_R - E_Q) \qquad (3)$$

Values of $E_B$ plotted in Fig. 13 show that an increase from 224 to 243 meV when the Ga content increases from 0 to 0.1. The same type of variation of $E_B$ with Ga doping was also reported by Sun et al [21], however their values of $E_B$ are about 70 meV lower for x = 0 and almost coincide for x = 0.1. Such a difference may be at least partially due to a different microstructure following a somewhat different heat treatment during the sample preparation.

In conclusion, the adiabatic polaron model seems to be of interest for explaining features seen in electrical transport properties. It takes a step to decide that the features observed on thermal transport do reveal the same mechanisms. The additional perspective of assisted tunnelling due to the disordered environment imposed by Ga substitution seems to smoothen any apparent disagreement, if any occurs. This is particularly the case for ''large'' Ga content substitution. Indeed (and in fact) Ga doping on Mn sites remarkably affects the temperature variation of TEP and a systematic set of behaviors – confirming a role played by Ga doping in localization phenomena. However TEP is a complex quantity to analyse, since it mixes entropy and scattering effects. A theory of TEP in substituted CMR materials seems regrettably lacking. The merging of TEP curves at high temperature might be of interest in order to understand the interplay between electronic density of states variation and spin scattering mechanisms by various excitations. It might be a crucial theoretical test. The same can be said of the thermal conductivity, - which we recall shows a multipeak structure, and its subsequent magnetic field dependence. Beyond the lowest temperatures the applied magnetic field enhances thermal conductivity in the manganites studied. One could also expect that the thermal conductivity should diminish with raising Ga doping, which destroys the double exchange paths. Moreover the Ga point defects act as additional scattering centers. Unfortunately the 10 % inaccuracy in geometrical dimensions of samples does not allow to directly compare a magnitude of thermal conductivity for both samples and to answer this question. The magnetoresistance effect is found to increase both with raising Ga doping and magnetic field strength. It is worth to notice that the higher Ga doping broadens the temperature interval as well as enhances the magnetoresistance magnitude (Fig. 9).

Ackn. Work supported in parts by PST.MEM.CLG.980654, Polish – Belgian Scientific Exchange Program, Scientific Net "Oxide materials with highly correlated electrons" and FNRS (Belgium) fellowship (BV). The help by Ph. Vanderbemden and J.F. Fagnard is kindly ackowledged.

**FIGURE CAPTIONS**

Fig.1. Magnetic moment per formula unit at 10 K as a function of the DC magnetic field for samples with different Ga content. The magnetic moment values are given in $\mu_B$/formula unit. Inset: Magnetic moment at 10 K and 3 T as a function of the gallium content (x). The dashed line corresponds to the theoretical spin-only values.

Fig.2. Temperature dependence of the in-phase component of the AC magnetic susceptibility for $\mu_0 H_{AC} = 1$ mT and f = 1053 Hz. The out-of-phase component of susceptibility for the sample with x = 0.10 is also shown (axis on the right).

Fig.3. Temperature dependence of the DC magnetization of the sample with x = 0.10, for $\mu_0 H_{DC} = 10$ mT (left axis) and 100 mT (right axis) under ZFC and FC conditions.

Fig. 4. Temperature dependence of normalized electrical resistance in zero magnetic field of samples $La_{0.7}Ca_{0.3}Mn_{1-x}Ga_xO_3$, x = 0, 0.025, 0.05, 0.1.

Fig. 5a-d. Temperature dependence of electrical resistance of samples $La_{0.7}Ca_{0.3}Mn_{1-x}Ga_xO_3$, x = 0, 0.025, 0.05, 0.1.

Fig. 6. Temperature position of resistivity maximum versus magnetic field for samples $La_{0.7}Ca_{0.3}Mn_{1-x}Ga_xO_3$, x = 0, 0.025, 0.05, 0.1.

Fig. 7. Magnetic field dependence of exponent N derived from electrical resistance for samples $La_{0.7}Ca_{0.3}Mn_{1-x}Ga_xO_3$, x = 0, 0.025, 0.05, 0.1.

Fig. 8. Magnetic field dependence of activation energy $E_R$ derived from electrical resistance for samples

Fig. 9a-d. Temperature dependence of magneto - resistance for samples $La_{0.7}Ca_{0.3}Mn_{1-x}Ga_xO_3$, x = 0, 0.025, 0.05, 0.1.

Fig. 10. Temperature variation of thermoelectric power of samples $La_{0.7}Ca_{0.3}Mn_{1-x}Ga_xO_3$, x = 0, 0.025, 0.05, 0.1.

Fig. 11. Temperature variation of thermal conductivity for $La_{0.7}Ca_{0.3}MnO_3$ at zero and 8 T fields and for $La_{0.7}Ca_{0.3}Mn_{0.95}Ga_{0.05}O_3$ at zero and 4 T fields.

Fig. 12. Temperature variation of a difference in thermal conductivity measured in the zero and high magnetic fields for $La_{0.7}Ca_{0.3}MnO_3$ and for $La_{0.7}Ca_{0.3}Mn_{0.95}Ga_{0.05}O_3$.

Fig. 13. Variation of polaron binding energy on Ga content x for samples $La_{0.7}Ca_{0.3}Mn_{1-x}Ga_xO_3$, x = 0, 0.025, 0.05, 0.1.

Table 1 : Magneto – transport parameters for $La_{0.7}Ca_{0.3}Mn_{1-x}Ga_xO_3$ manganites.

| sample | x | Curie temperature | Magnetic moment / f.u. | Electrical resistivity at 10 K | $R_{RT}$ / R(10K) | TEP break temperature $T_B$ | Temperature of TEP minimum |
|---|---|---|---|---|---|---|---|
| | | K | $\mu_B$ | $\Omega m$ | | K | K |
| | | | | | | | |
| $La_{0.7}Ca_{0.3}MnO_3$ | 0 | 249 | 3.71 | $2 \times 10^{-5}$ | 18 | - | 189 |
| $La_{0.7}Ca_{0.3}Mn_{0.975}Ga_{0.025}O_3$ | .025 | 219 | 3.58 | $9 \times 10^{-5}$ | 22 | 210 | 227 |
| $La_{0.7}Ca_{0.3}Mn_{0.95}Ga_{0.05}O_3$ | .05 | 168 | 3.49 | $1 \times 10^{-4}$ | 57 | 158 | 188 |
| $La_{0.7}Ca_{0.3}Mn_{0.90}Ga_{0.10}O_3$ | .1 | 122 | 3.29 | $3 \times 10^{-1}$ | 16 | 70 | 143 |

Tab. 2. Correlation coefficient R for electrical resistivity fitted to adiabatic polaron model in temperature intervals between $T_P$ and 340 K for $La_{0.7}Ca_{0.3}Mn_{1-x}Ga_xO_3$ manganites.

| Magnetic field (T) | x = 0 | | x = 0.025 | | x = 0.05 | | x = 0.1 | |
|---|---|---|---|---|---|---|---|---|
| | R | $T_P$ (K) | R | $T_P$ (K) | R | $T_P$ (K) | R | $T_P$ (K) |
| 0 | .99987 | 267 | .99982 | 233 | .99983 | 189 | .99847 | 143 |
| .2 | .99997 | 272 | .99985 | 238 | .99978 | 192 | .99964 | 168 |
| .5 | .99978 | 271 | .99981 | 238 | .99965 | 204 | .99950 | 167 |
| 1 | .99972 | 277 | .99968 | 244 | .99954 | 200 | .99924 | 167 |
| 2 | .99941 | 298 | .99979 | 260 | .9998 | 223 | .99971 | 201 |
| 4 | .99803 | 308 | .99959 | 287 | .99979 | 250 | .99965 | 222 |
| 6 | .99646 | 312 | .99947 | 300 | .99829 | 251 | .99968 | 227 |
| 8 | .99156 | 322 | .99879 | 312 | .99982 | 280 | .99973 | 233 |

*FIGURE CAPTIONS*

*Fig.1. Magnetic moment per formula unit at 10 K as a function of the DC magnetic field for samples with different Ga content. The magnetic moment values are given in $\mu_B$/formula unit. Inset: Magnetic moment at 10*

K and 3 T as a function of the gallium content (x). The dashed line corresponds to the theoretical spin-only values. **Gallium~1- bv**

Fig.2. Temperature dependence of the in-phase component of the AC magnetic susceptibility for $\mu_0 H_{AC} = 1$ mT and $f = 1053$ Hz. The out-of-phase component of susceptibility for the sample with $x = 0.10$ is also shown (axis on the right). **Gallium~1- bv**

Fig.3. Temperature dependence of the DC magnetization of the sample with $x = 0.10$, for $\mu_0 H_{DC} = 10$ mT (axis on the left) and 100 mT (axis on the right) under ZFC and FC conditions. **Gallium~1- bv**

Fig. 4. Temperature dependence of normalized electrical resistance in zero magnetic field of samples $La_{0.7}Ca_{0.3}Mn_{1-x}Ga_xO_3$, $x = 0, 0.025, 0.05, 0.1$. **GaBV-RN**

**Fig. 5a-d. Temperature dependence of electrical resistance of samples $La_{0.7}Ca_{0.3}Mn_{1-x}Ga_xO_3$, $x = 0, 0.025, 0.05, 0.1$.** **4x**

Fig. 6. Temperature position of resistivity maximum versus magnetic field for samples $La_{0.7}Ca_{0.3}Mn_{1-x}Ga_xO_3$, $x = 0, 0.025, 0.05, 0.1$. **. bv-P**

Fig. 7. Magnetic field dependence of exponent N derived from electrical resistance for samples $La_{0.7}Ca_{0.3}Mn_{1-x}Ga_xO_3$, $x = 0, 0.025, 0.05, 0.1$. **bv-N**

Fig. 8. Magnetic field dependence of activation energy $E_R$ derived from electrical resistance for samples $La_{0.7}Ca_{0.3}Mn_{1-x}Ga_xO_3$, $x = 0, 0.025, 0.05, 0.1$. **bv-ER-2**

Fig. 9a-d. Temperature dependence of magneto - resistance for samples $La_{0.7}Ca_{0.3}Mn_{1-x}Ga_xO_3$, $x = 0, 0.025, 0.05, 0.1$. **4 x La7Ca3-m**

Fig. 10. Temperature variation of thermoelectric power of samples $La_{0.7}Ca_{0.3}Mn_{1-x}Ga_xO_3$, $x = 0, 0.025, 0.05, 0.1$. **Ga-bv-Q2**

Fig. 11. Temperature variation of thermal conductivity for $La_{0.7}Ca_{0.3}MnO_3$ at zero and 8 T fields and for $La_{0.7}Ca_{0.3}Mn_{0.95}Ga_{0.05}O_3$ at zero and 4 T fields. **Bvk + Ga-bv-k2**

Fig. 12. Temperature variation of a difference in thermal conductivity measured in the zero and high magnetic fields for $La_{0.7}Ca_{0.3}MnO_3$ and for $La_{0.7}Ca_{0.3}Mn_{0.95}Ga_{0.05}O_3$. **BVkD**

**Fig. 13. Variation of polaron binding energy on Ga content x for samples $La_{0.7}Ca_{0.3}Mn_{1-x}Ga_xO_3$, $x = 0, 0.025, 0.05, 0.1$.** **EB**

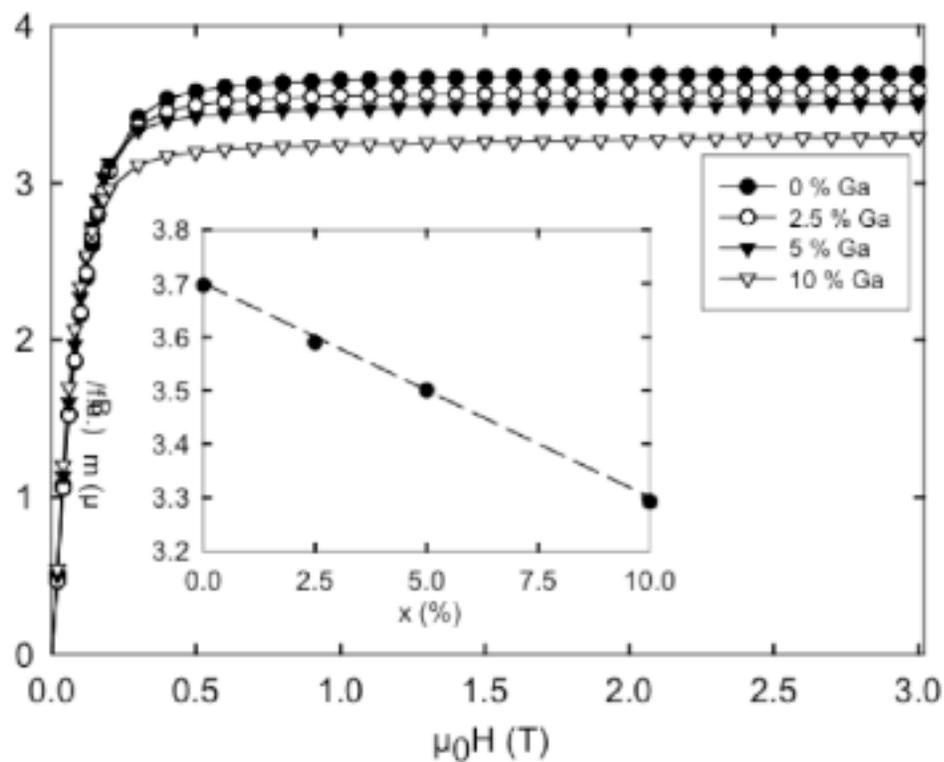
Fig. 1

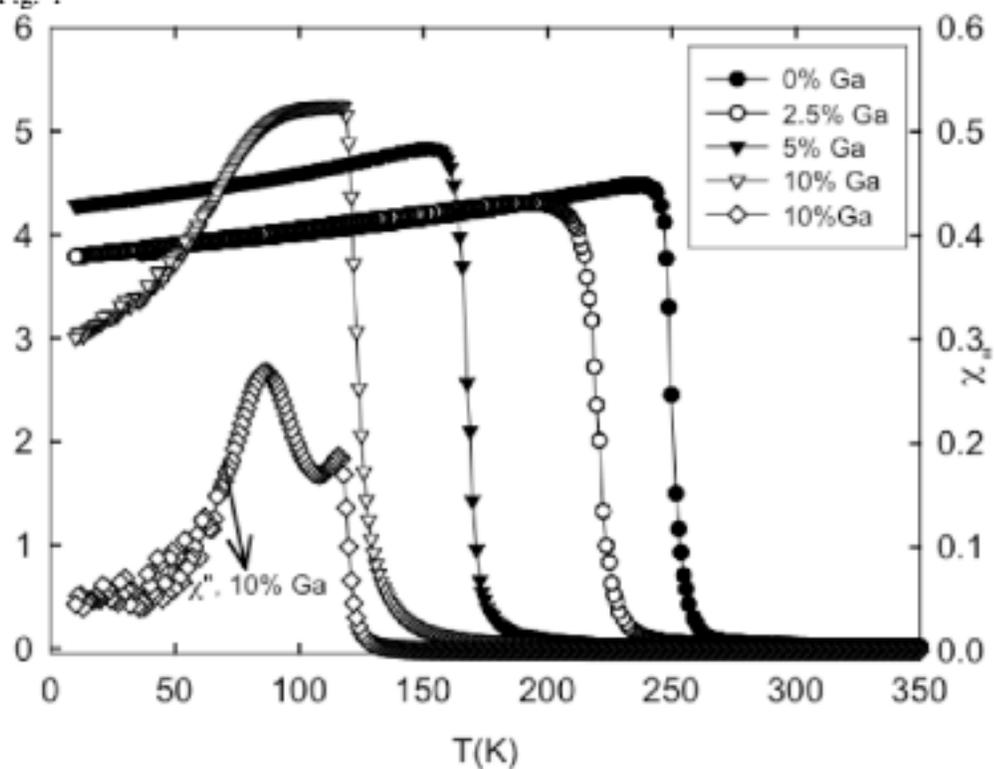
Fig. 2

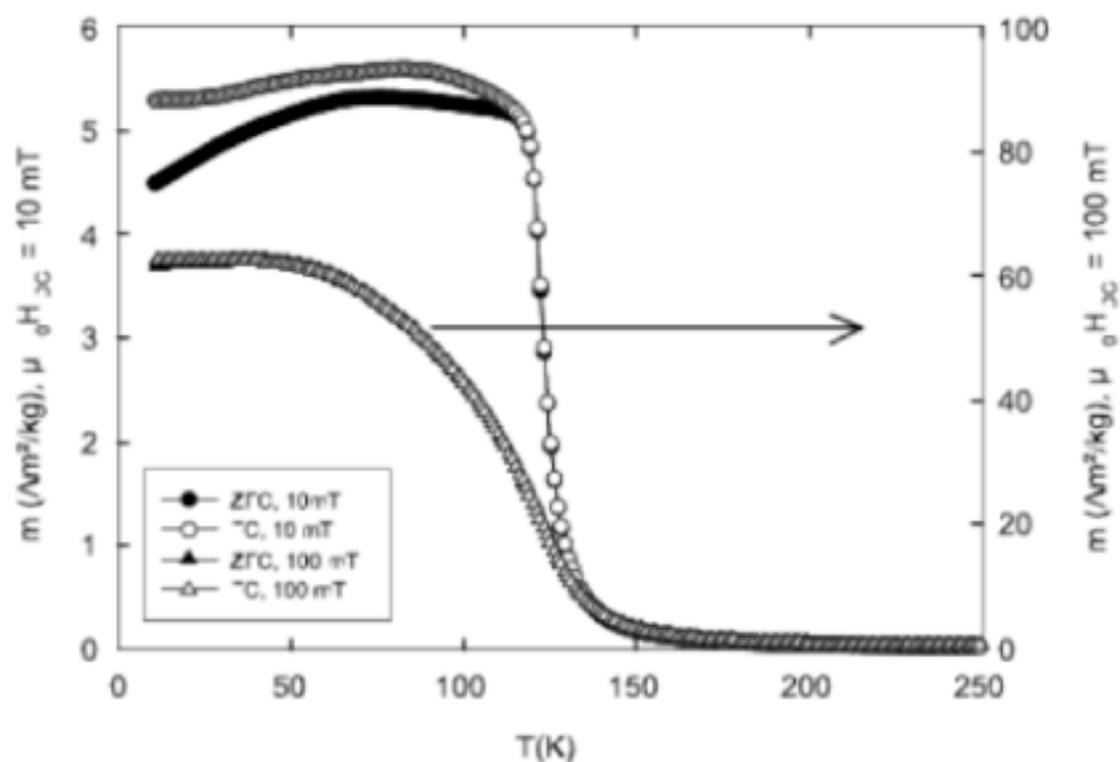

Fig.3

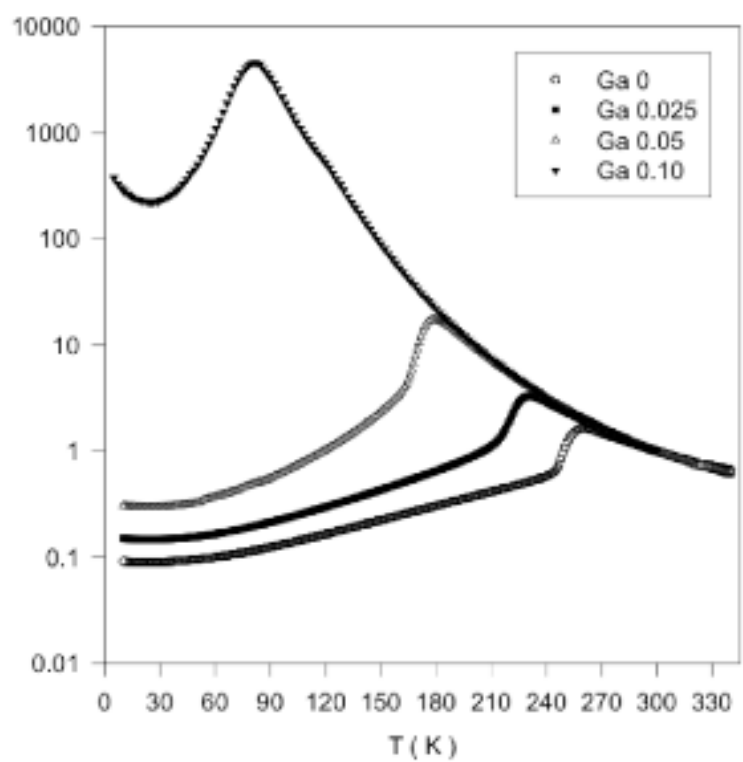

Fig. 4

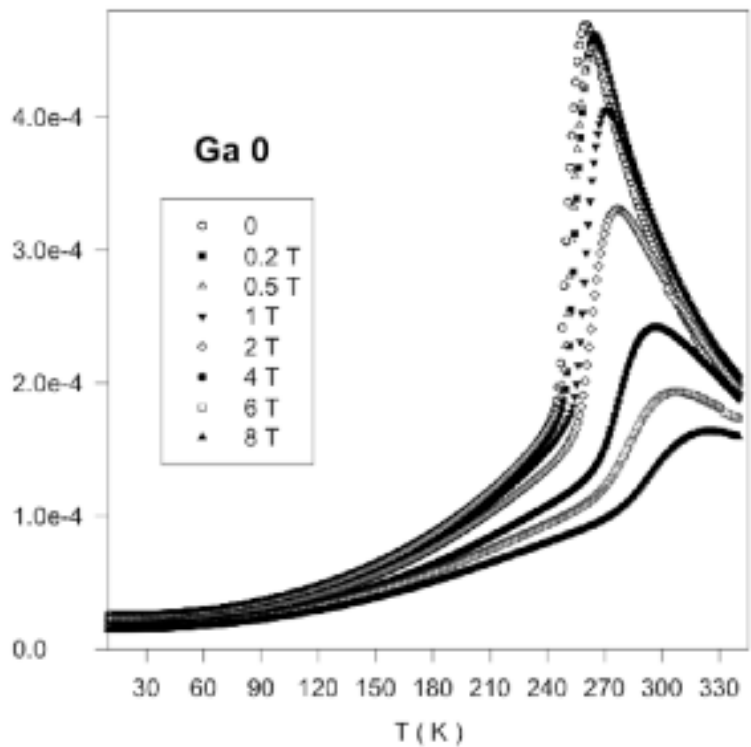
Fig. 5A
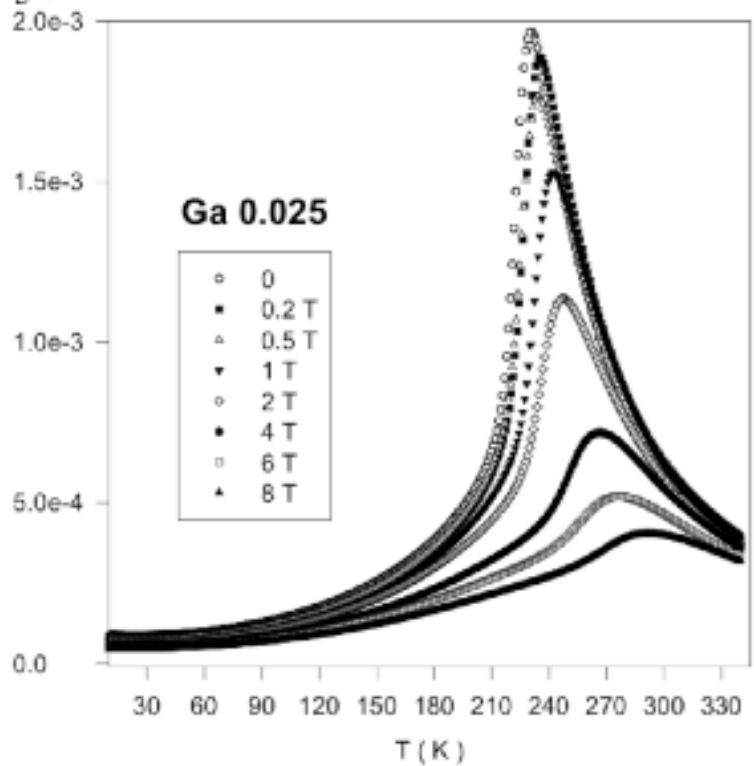
Fig. 5B

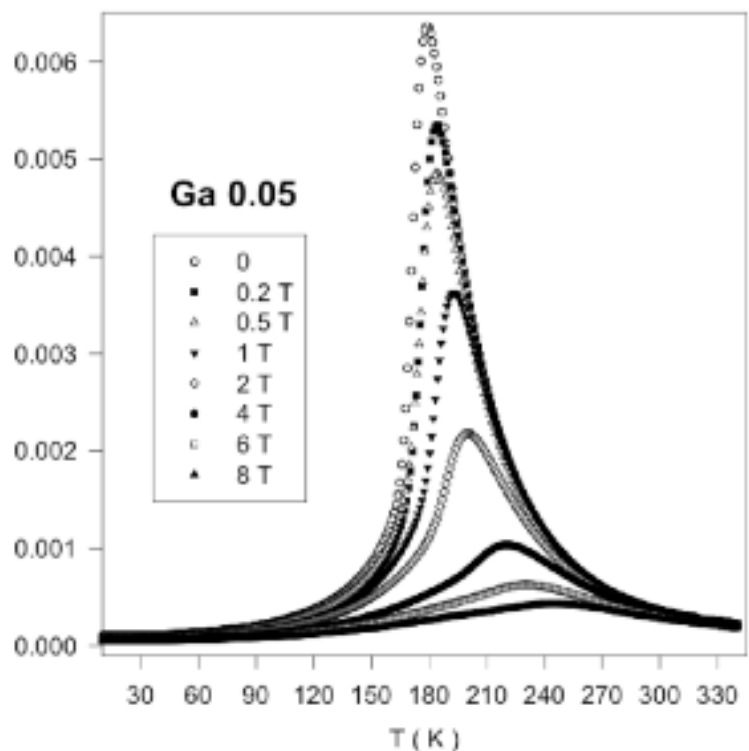
Fig. 5C

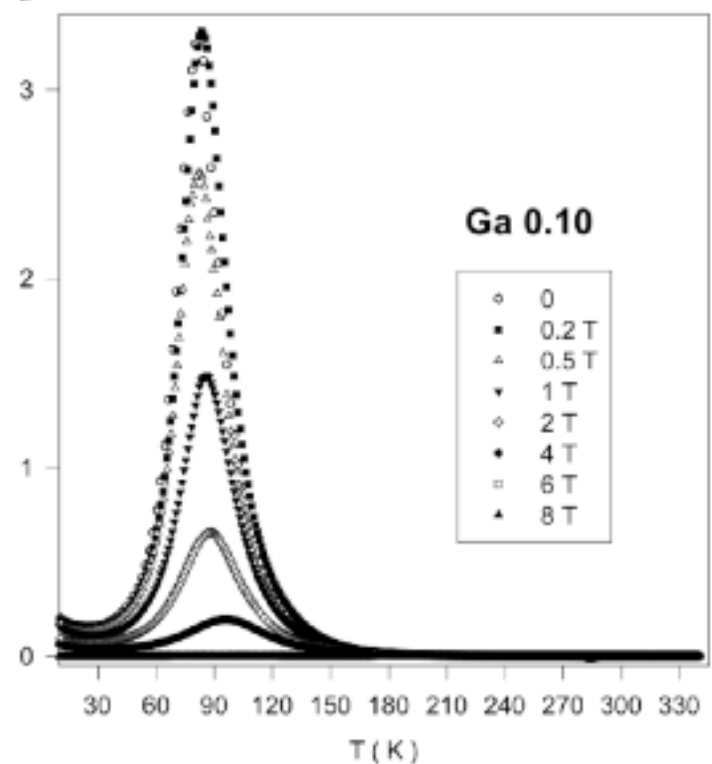
Fig. 5D

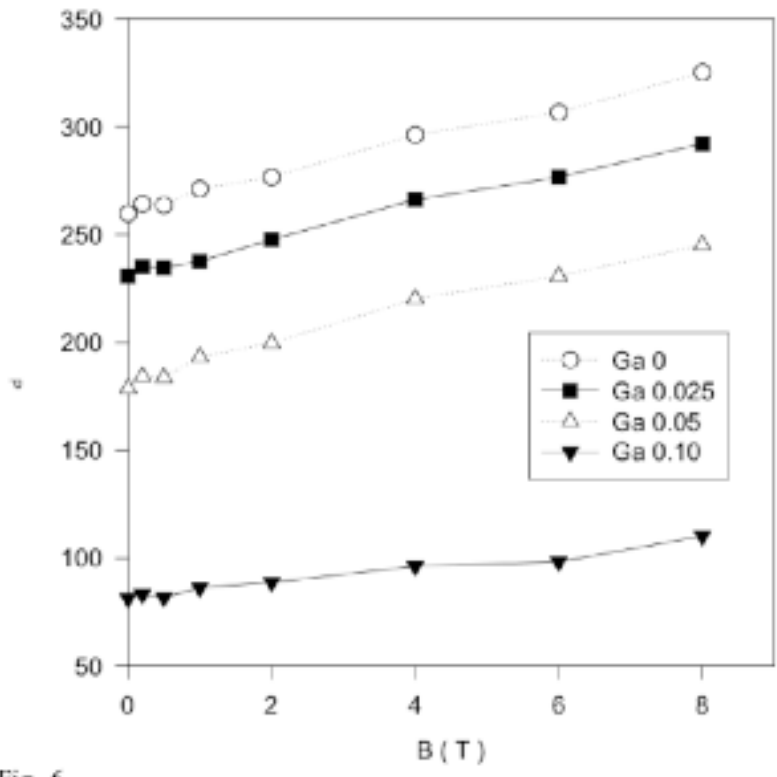

Fig. 6

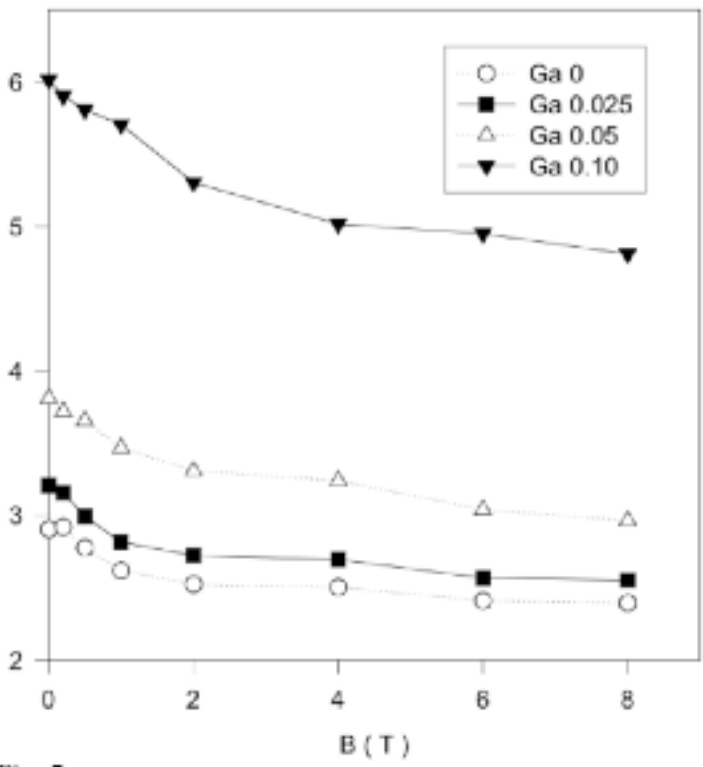

Fig. 7

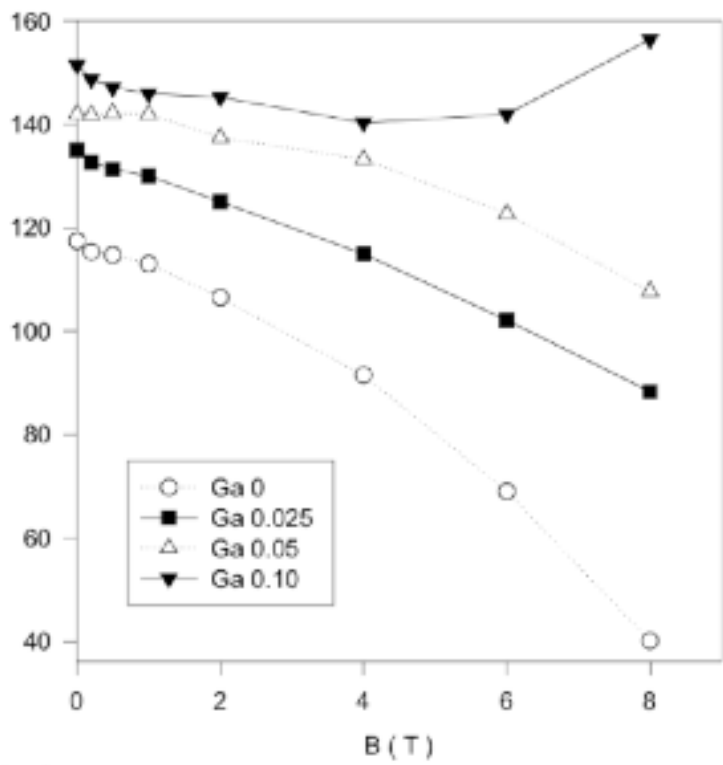

Fig. 8

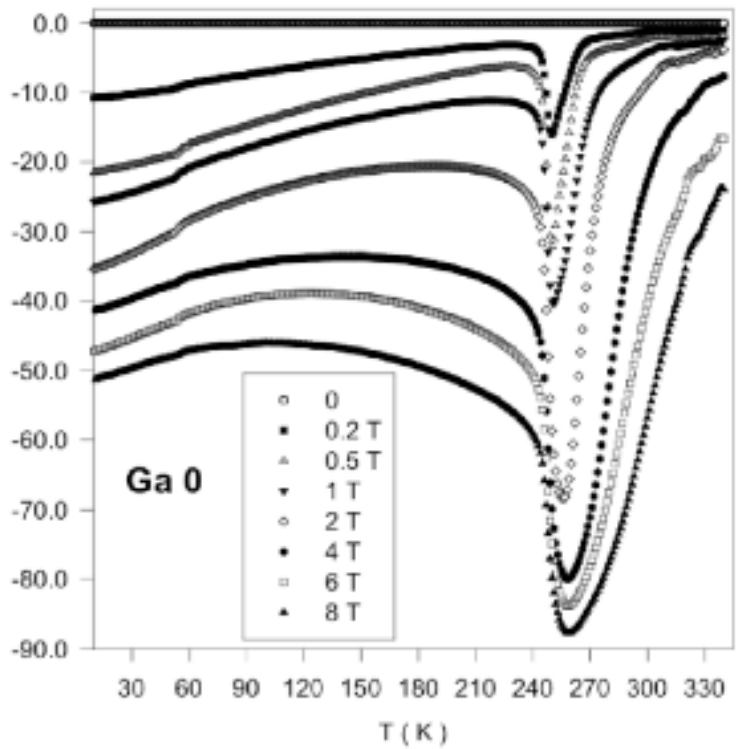

Fig. 9A

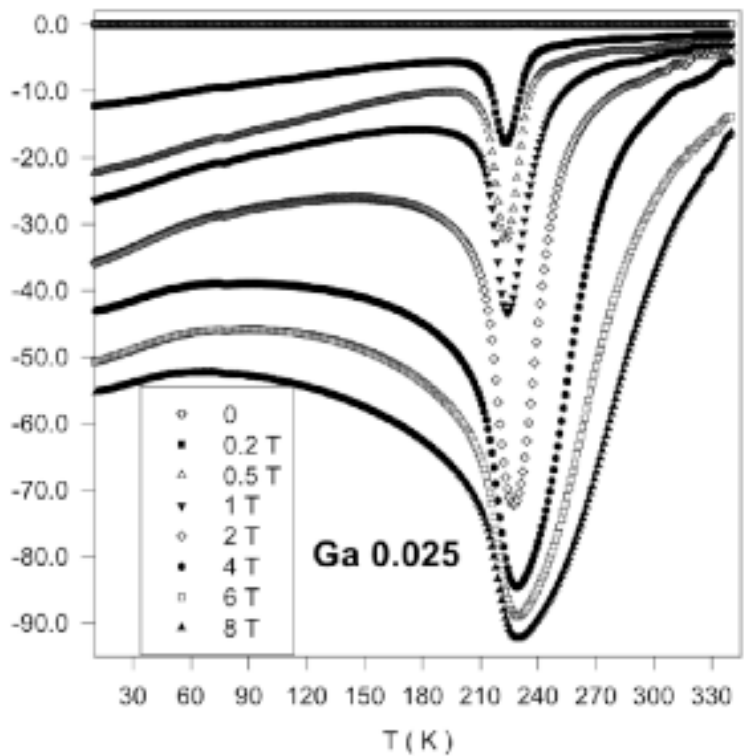
Fig. 9B
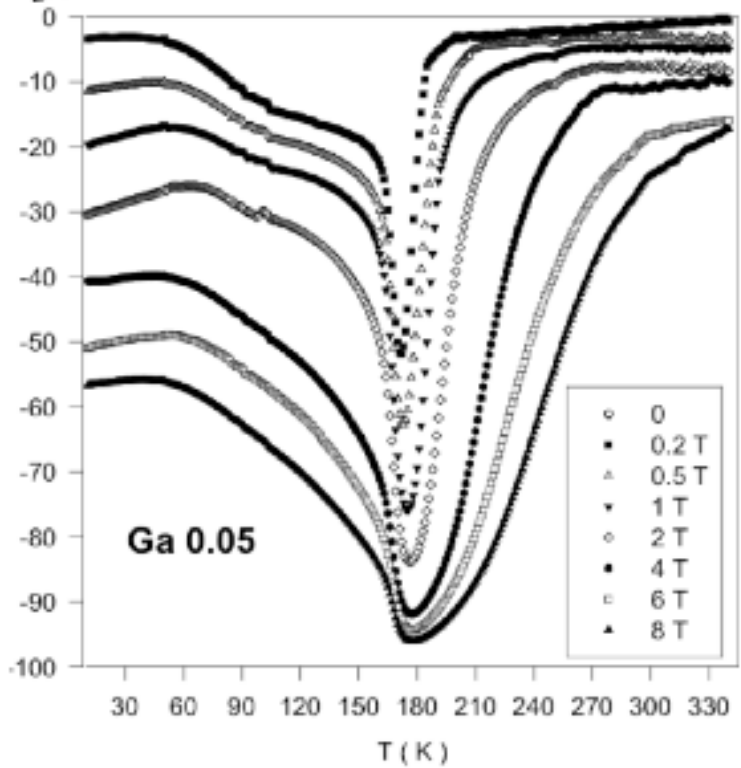
Fig. 9C

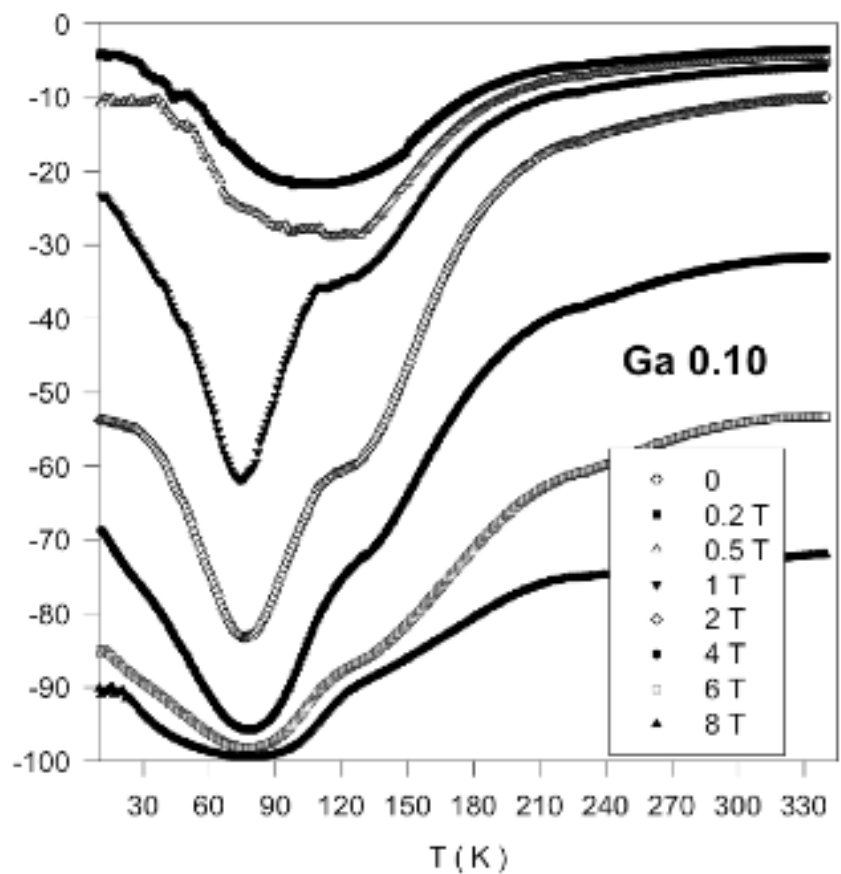
Fig. 9D

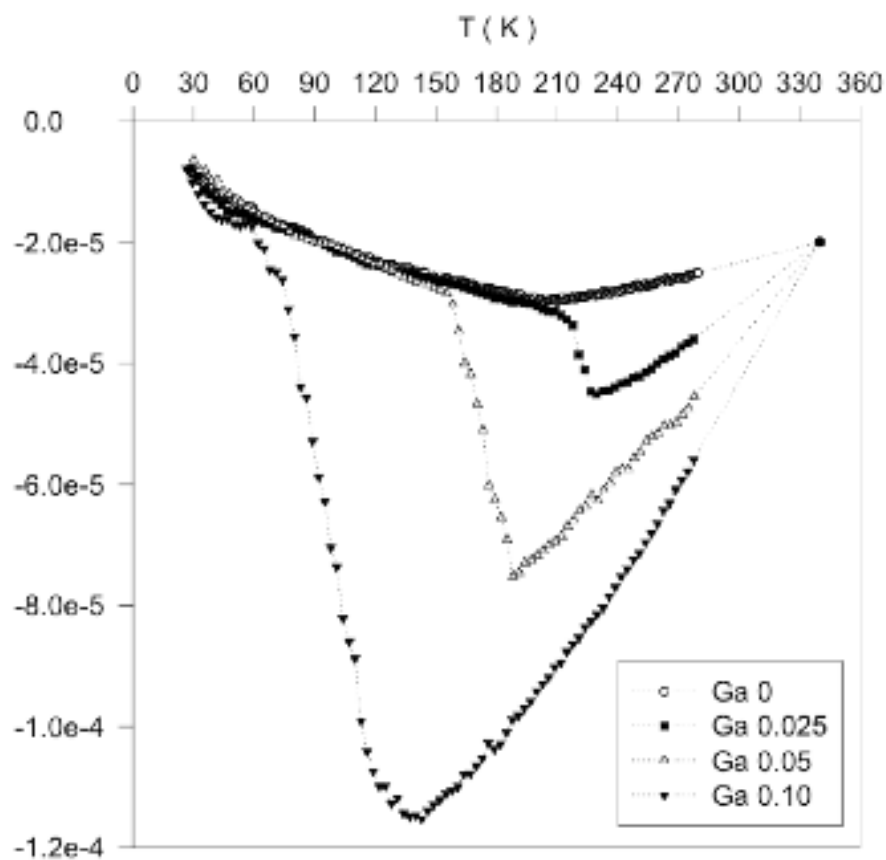
Fig. 10